\documentclass[pra,twocolumn,superscriptaddress,amsmath,amssymb,floatfix]{revtex4}
\usepackage{graphicx,pstricks,subfigure}
\usepackage{epstopdf}
\usepackage{bm}

\newcommand{\Op}[1]{\boldsymbol{\mathsf{\hat{#1}}}}
\def\openone{\leavevmode\hbox{\small1\kern-3.3pt\normalsize1}}

\begin{document}

\author{Subhas Ghosal}
\email{E-mail: subhas.ghosal@yahoo.co.in}

\affiliation{Department of Chemistry, Durham University, South
Road, DH1 3LE, United Kingdom}

\author{Richard J. Doyle}

\affiliation{Department of Chemistry, Durham University, South
Road, DH1 3LE, United Kingdom}

\author{Christiane P. Koch}
\email{E-mail: ckoch@physik.fu-berlin.de}

\affiliation{Institut f\"{u}r Theoretische Physik, Freie
Universit\"{a}t Berlin, Arnimallee 14, 14195 Berlin, Germany}

\author{Jeremy M. Hutson}
\email{E-mail: J.M.Hutson@durham.ac.uk}

\affiliation{Department of Chemistry, Durham University, South
Road, DH1 3LE, United Kingdom}

\title{Stimulating the production of deeply bound RbCs molecules
with laser pulses: \\ the role of spin-orbit coupling in
forming ultracold molecules}

\date{\today}

\begin{abstract}
We investigate the possibility of forming deeply bound
ultracold RbCs molecules by a two-color photoassociation
experiment. We compare the results with those for Rb$_2$ in
order to understand the characteristic differences between
heteronuclear and homonuclear molecules. The major differences
arise from the different long-range potential for excited
states. Ultracold $^{85}$Rb and $^{133}$Cs atoms colliding on
the X\,$^1\Sigma^+$ potential curve are initially
photoassociated to form excited RbCs molecules in the region
below the Rb(5S) + Cs(6P$_{1/2}$) asymptote. We explore the
nature of the $\Omega=0^+$ levels in this region, which have
mixed A\,$^1\Sigma^+$ and b\,$^3\Pi$ character. We then study
the quantum dynamics of RbCs by a time-dependent wavepacket
(TDWP) approach. A wavepacket is formed by exciting a few
vibronic levels and is allowed to propagate on the coupled
electronic potential energy curves. We calculate the
time-dependence of the overlap between the wavepacket and
ground-state vibrational levels. For a detuning of 7.5
cm$^{-1}$ from the atomic line, the wavepacket for RbCs reaches
the short-range region in about 13 ps, which is significantly
faster than for the homonuclear Rb$_2$ system; this is mostly
because of the absence of an $R^{-3}$ long-range tail in the
excited-state potential curves for heteronuclear systems. We
give a simple semiclassical formula that relates the time taken
to the long-range potential parameters. For RbCs, in contrast
to Rb$_2$, the excited-state wavepacket shows a substantial
peak in singlet density near the inner turning point, and this
produces a significant probability of deexcitation to form
ground-state molecules bound by up to 1500 cm$^{-1}$. The
short-range peak depends strongly on nonadiabatic coupling and
is reduced if the strength of the spin-orbit coupling is
increased. Our analysis of the role of spin-orbit coupling
concerns the character of the mixed states in general and is
important for both photoassociation and stimulated Raman
deexcitation.
\end{abstract}

\pacs{32.80.Qk, 33.80.Ps, 34.50.Rk}

\maketitle

\section{Introduction}

The study of cold molecules, below 1 K, and ultracold
molecules, below 1 mK, offers many new opportunities for
chemical and molecular physics. Cold molecules open up a new
regime for molecular collisions in which classical physics
breaks down completely and a quantal description is needed for
all degrees of freedom. Collisions in this regime are dominated
by long-range forces and exhibit resonance phenomena that can
be controlled by applied electric and magnetic fields. Cold
molecules also present new opportunities for precision
measurement and open up possibilities for measuring quantities
such as the dipole moment of the electron and the
time-dependence of fundamental ``constants''.

There is particular interest in ultracold polar molecules,
because dipolar species interact more strongly and at much
longer range than non-polar species. Dipolar quantum gases are
predicted to exhibit new phenomena such as anisotropic
Bose-Einstein condensation \cite{Leggett:2001} and may have
applications in quantum information processing
\cite{RablPRL06}. They also provide opportunities for
engineering highly correlated quantum phases
\cite{BuechlerPRL07}.

Molecules can be formed in ultracold atomic gases by both
magnetoassociation and photoassociation \cite{Jones:RMP:2006,
Hutson:IRPC:2006, Kohler:RMP:2006}, and there have been
considerable advances in using such methods to produce
heteronuclear alkali metal dimers \cite{Wang:KRb-PRL:2004,
Sage:2005, Hudson:PRL:2008, Kerman:vibdistRbCs:2004,
Kerman:paRbCs:2004, Haimberger:2004, Kraft:JPB:2006,
Deiglmayr:2008}. However, both photo- and magnetoassociation
initially produce molecules in very high vibrational levels,
which are only very weakly dipolar even for heteronuclear
species. There is therefore great interest in producing
ultracold molecules in low-lying vibrational states. For
example, Sage \emph{et al.}\ \cite{Sage:2005, Hudson:PRL:2008}
have succeeded in producing small numbers of ultracold RbCs
molecules in the ground vibronic state by a four-photon
photoassociation scheme using continuous-wave lasers, while
Deiglmayr \emph{et al.}\ \cite{Deiglmayr:2008} have produced
LiCs molecules with a two-photon scheme. Both these approaches
include a spontaneous emission step, but Ni \emph{et al.}\
\cite{Ni:KRb:2008} have very recently produced KRb molecules in
the lowest vibrational levels of the lowest singlet and triplet
electronic states by magnetoassociation followed by a coherent
two-photon process (stimulated Raman adiabatic passage,
STIRAP). Danzl \emph{et al.}\ \cite{Danzl:ground:2008} have
carried out analogous experiments on the homonuclear molecule
Cs$_2$, while Lang {\em et al.} \cite{Lang:ground:2008} have
produced Rb$_2$ molecules in the lowest level of the lowest
triplet state.

Heteronuclear molecules differ from homonuclear molecules in
several ways. The most important difference is that the
excited-state potential curves correlating with $^2$S + $^2$P
atoms have an $R^{-6}$ behavior at long range in the
heteronuclear case but an $R^{-3}$ behavior in the homonuclear
case, due to the resonant dipole interaction. Because of this,
the Franck-Condon factors for photoassociation are quite
different. In addition, the density of vibrational levels in
the electronically excited state is different for $1/R^{6}$ and
$1/R^{3}$ potentials.

Magnetoassociation must be carried out in tight traps at very
low temperatures. However, there would be advantages in
producing molecules in low-lying states at the somewhat higher
temperatures (in the microkelvin regime) that are available in
magnetooptical traps (MOTs), where the number of atoms
available is often much larger. A very promising approach for
this is photoassociation using shaped laser pulses. Short-lived
molecules are formed in the excited electronic state by
photoassociation using a short laser pulse (pump pulse) during
the collision of two ultracold atoms. The excited molecules are
then stabilized by stimulated emission using a second laser
pulse (dump pulse) into the bound vibrational levels of the
ground electronic state. Luc-Koenig {\em et al.}
\cite{Luc-Koenig:PRA:2004,Luc-Koenig:epd:2004} and Koch {\em et
al.} \cite{Koch:PRA:2006a,Koch:PRA:2006b} have simulated this
process for homonuclear diatomic molecules such as Cs$_2$ and
Rb$_2$, and initial experiments on these systems have been
reported by Salzmann \emph{et al.}\ \cite{Salzmann:2006,
Salzmann:2008} and Brown \emph{et al.}\ \cite{Brown:2006}.
Similar experiments can be envisioned for heteronuclear
molecules, with the use of evolutionary algorithms or other
strategies to maximize the rate of production of ground-state
molecules. However, a theoretical study of short-pulse
photoassociation of heteronuclear molecules has not yet been
carried out. In this paper, we consider RbCs as a prototype
heteronuclear molecule and we study the possibility of forming
ground-state RbCs molecules through pump-dump photoassociation.

\begin{figure}[tbp]
\includegraphics*[width=0.9\linewidth]{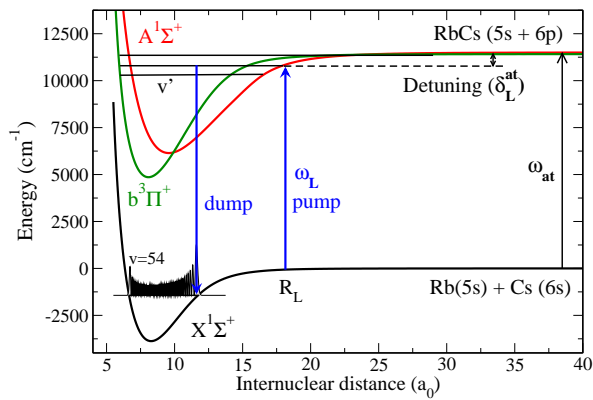}
\caption{The potential curves and laser pulse scheme considered
here. The potential curves correspond to the ground
X\,$^1\Sigma^+$ and the excited A\,$^1\Sigma^+$ and b\,$^3\Pi$
states. A picosecond laser pulse is used to excite few vibronic
levels in the excited state to form a wavepacket. A second
laser pulse is used to dump the wavepacket to the ground
X\,$^1\Sigma^+$ state after an appropriate time delay.
The vibrational wavefunction of the target level of the dump pulse,
$v=54$, is also displayed.
\label{scheme}}
\end{figure}

We consider the photoassociation of a colliding pair of
$^{85}$Rb and $^{133}$Cs atoms, initially in their ground
state. Absorption of a photon red-detuned from the atomic
Rb(5S) + Cs(6P$_{1/2}$) asymptote forms an excited-state
molecule around internuclear distance $R_L$, as shown in Fig.\
\ref{scheme}. When spin-orbit coupling is included there are 8
electronic states that correlate with the excited 5S$_{1/2}$ +
6P$_{1/2}$ and 5S$_{1/2}$ + 6P$_{3/2}$ asymptotes: two $0^+$
states, two $0^-$ states, three $\Omega=1$ states and one
$\Omega=2$ state \cite{Kotochigova:2005}. These correlate at
short range with the A\,$^1\Sigma^+$, b\,$^3\Pi$, B\,$^1\Pi$
and c\,$^3\Sigma^+$ electronic states in Hund's case (a)
labelling.

Absorption of a photon from the ground X\,$^1\Sigma^+$ ($0^+$)
state can produce excited-state molecules in $0^+$ or 1 states,
while absorption from the a\,$^3\Sigma^+$ state (with $0^-$ and
1 components) can produce $0^+$, $0^-$, 1 and 2 states. We
focus here on the $0^+$ excited states, which may be formed by
combining the A\,$^1\Sigma^+$ and b\,$^3\Pi$ states. The
c\,$^3\Sigma^+$ state does not contribute because it has only
$0^-$ and 1 components and the B\,$^1\Pi$ state does not
contribute because it has only an $\Omega=1$ component.
Rotational couplings that would connect states of different
$\Omega$ are neglected. This 2-state approximation is analogous
to the approach used for the $0_u^+$ states in Cs$_2$ and
Rb$_2$ \cite{kokoouline:JCP:1999}, for which the strongly mixed
singlet-triplet character of the vibronic wavefunctions was
found to produce enhanced formation of ground-state molecules
with binding energies on the order of 10~cm$^{-1}$
\cite{Koch:PRA:2006b, Pechkis:PRA:2007}.

The pump pulse produces a non-stationary state made up of
long-range excited-state levels of mixed singlet and triplet
character. The corresponding wavepacket propagates towards
short range under the influence of the excited-state
potentials. After a suitable time delay, when a sufficient
amount of the wavepacket has reached the short-range region, a
dump pulse is activated to transfer the molecules into
vibrational levels of the X\,$^1\Sigma^+$ electronic ground
state.

We simulate the entire pump-dump process using a time-dependent
wavepacket (TDWP) approach. Our goal is to optimize the
parameters of the pump and dump pulses to maximize the
production of molecules in deeply-bound levels of the ground
state. We find that a crucial factor is the magnitude of the
spin-orbit coupling near the avoided crossing. If the
spin-orbit coupling is sufficiently strong, then the levels
that lie on the lower adiabatic curve, which correlates with 5S
+ 6P$_{1/2}$ at long range, have mainly triplet character at
short range. Such levels have low intensities for deexcitation.
We therefore form the excited-state wavepacket by selective
photoassociation into vibronic levels of the A\,$^1\Sigma^+$-
b\,$^3\Pi$ electronically excited states that are strongly
mixed by nonadiabatic coupling and have enhanced singlet
character at short range. We estimate the parameters required
for the dump pulse and the time delay by analyzing the
time-dependence of the overlap between the singlet part of the
wavepacket and the vibrational levels of the ground electronic
state.

We find an important difference between photoassociation for
heteronuclear and homonuclear molecules. The excited states of
heteronuclear molecules have a lower density of vibrational
levels near dissociation than those of homonuclear molecules,
because of the long-range $R^{-3}$ term in the excited-state
potentials for homonuclear species. Because of this, the atoms
of heteronuclear molecules produced at long range experience a
significantly faster classical acceleration towards one another
than homonuclear molecules created at the same binding energy.
The time delay required between the pump and dump pulses is
thus smaller in the heteronuclear case (about 13 ps for RbCs
with a detuning of 7.5 cm$^{-1}$ from the atomic line) than in
the homonuclear case. However, even for heteronuclear
molecules, there is little probability of producing
ground-state molecules within the duration of a single pulse of
less than a few ps.

The paper is organized as follows. In section II, we discuss
the potential energy curves and spin-orbit coupling functions
used in this study. In sections III and IV we explore the
dependence of the zero-energy $s$-wave scattering lengths on
the interaction potentials and discuss their influence on the
Franck-Condon factors for photoassociation. In section V we
describe wavepacket studies of the photoassociation process and
the subsequent formation of  ground-state molecules. At each
stage, the results for RbCs are compared with those for Rb$_2$
in order to establish the similarities and differences between
heteronuclear and homonuclear molecules.

\section{Potential Curves}
Several studies of the potential energy curves for RbCs have
been published. Obtaining accurate potentials is not a
straightforward task, because electronic structure calculations
are very difficult for such heavy atoms. Experimental results
are available for some states \cite{Fellows:1999,
Bergeman:2003}, but are sparse or nonexistent for others. Only
some of the theoretical potentials include spin-orbit coupling
effects; for example the relativistic curves of Kotochigova and
Tiesinga \cite{Kotochigova:2005} include avoided crossings
which are absent in the older pseudopotential results of
Allouche \emph{et al}. \cite{Allouche:2000}.

The potential curves for the electronic states of RbCs
considered here, neglecting spin-orbit coupling, are shown in
Fig.\ \ref{scheme}. From the viewpoint of photoassociation, the
most important parts of the curves are the long-range tails. In
this study we have chosen for simplicity to use the \emph{ab
initio} results of Allouche \emph{et al.}\ \cite{Allouche:2000}
at short range (bond lengths $R<15$ $a_0$). For the long-range
potentials, in the ground state, we use the experimentally
derived parameters of Fellows \emph{et al.}\
\cite{Fellows:1999}, with C$_6$ replaced by the more recent
value of Derevianko \emph{et al.}\ \cite{Derevianko:2001}.
These parameters are equivalent to column V of Table 2 in Ref.\
\onlinecite{Jamieson:2003}. The Le Roy radius for the ground X
$^1\Sigma^+$ state is 24.4 $a_0$, but a smooth match between
the long-range and short-range potentials was best achieved at
16.5 $a_0$. For the excited singlet and triplet states, we use
the theoretical long-range parameters of Marinescu and
Sadeghpour \cite{Marinescu:1999}, corresponding to the Rb(5S) +
Cs(6P) asymptote. For the excited states the matching between
long-range and short-range curves was achieved at 16.5 $a_0$
and 24 $a_0$ for the A and b states respectively.

\begin{figure}[tbp]
\includegraphics*[width=0.9\linewidth]{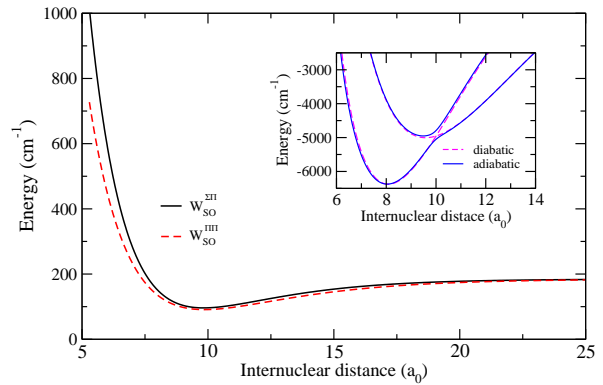}
\caption{The diagonal and off-diagonal components $W^{\Pi
\Pi}_{\rm SO}(R)$ and $W^{\Sigma \Pi}_{\rm SO}(R)$ of the
spin-orbit coupling as a function of $R$. The avoided crossing
between the two adiabatic potentials is shown in the inset.
\label{pes}}
\end{figure}

Spin-orbit coupling between the A\,$^1\Sigma^+$ and b\,$^3\Pi$
states plays an important role. Following Marinescu and
Dalgarno \cite{Marinescu:1996} and Aubert-Fr\'{e}con \emph{et
al}. \cite{Aubert:1998}, the effect of the spin-orbit coupling
for the $0^+$ states is included in terms of a 2$\times$2
electronic Hamiltonian matrix,
\begin{eqnarray}
H^{\rm el} =
\begin{pmatrix}V_{{\rm A}\,^1\Sigma^+}(R) & \sqrt{2}W_{\rm SO}^{\Sigma \Pi}(R) \\
\sqrt{2}W_{\rm SO}^{\Sigma \Pi}(R) & V_{{\rm b}\,^3\Pi}(R) - W_{\rm SO}^{\Pi \Pi}(R)
\end{pmatrix}, \nonumber \\
\label{ham_elec}
\end{eqnarray}
where $V_{{\rm A}\,^1\Sigma^+}(R)$ and $V_{{\rm b}\,^3\Pi}(R)$
are the Born-Oppenheimer potentials neglecting spin-orbit
coupling. The diagonal and off-diagonal spin-orbit coupling
elements are denoted $W_{\rm SO}^{\Pi \Pi}(R)$ and $W_{\rm
SO}^{\Sigma \Pi}(R)$ respectively and are functions of the
internuclear separation $R$ as shown in Fig.\ \ref{pes}. At
long range,
\begin{eqnarray}
W_{\rm SO}^{\Pi \Pi}(R \rightarrow \infty) =
W_{\rm SO}^{\Sigma \Pi}(R \rightarrow \infty) &=&
\frac{\Delta E_{\rm SO}^{\rm Cs}}{3}, \nonumber
\end{eqnarray}
where $\Delta E_{\rm SO}^{\rm Cs}$ is the spin-orbit splitting
of the $P$ state of atomic Cs. Bergeman {\em et al.}
\cite{Bergeman:2004} give a non-zero matrix element connecting
the $\Omega=0$ components of the b\,$^3\Pi$ and c\,$^3\Sigma^+$
states but this is appropriate only for the $0^-$ component and
does not influence the $0^+$ states. We use the spin-orbit
coupling functions obtained by Fellows and Bergeman
\cite{Fellows:priv:2007}, which were obtained by fitting to
results from Fourier-transform spectroscopy
\cite{Bergeman:2003}. The diagonal and off-diagonal spin-orbit
couplings show a significant dip from their asymptotic values
near the crossing between the A\,$^1\Sigma^+$ and b\,$^3\Pi$
potential energy curves. The two adiabatic potentials obtained
by diagonalizing the electronic Hamiltonian of Eq.\
(\ref{ham_elec}) are shown as an insert in Fig.\ \ref{pes}.

For Rb$_2$ the potentials are more accurately known. In the
present work, the ground-state X\,$^1\Sigma_g^+$ potential is
obtained by combining the short-range results of Seto \emph{et
al.}\ \cite{Seto:JCP:2000} with the long-range coefficients of
Marte \emph{et al.}\ \cite{Marte:2002}. The excited-state
potentials are taken from Bergeman \emph{et al.}\
\cite{Bergeman:2006}, who obtained potential curves and
spin-orbit coupling functions for the A\,$^1\Sigma_u^+$ and
b\,$^3\Pi_u$ states by combining photoassociation spectroscopy
with short-range {\em ab initio} results
\cite{Edvardsson:2003}. The corresponding curves for Rb$_2$
(not shown here) are qualitatively similar to those for RbCs,
except for the important difference in long-range behavior.

\begin{figure}[tbp]
\includegraphics*[width=\linewidth]{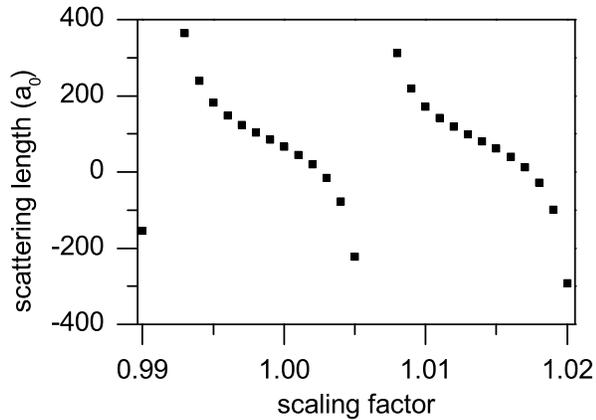}
\caption{RbCs s-wave scattering length for the X\,$^1\Sigma^+$ state,
as a function of a scaling factor used to multiply the potential.
\label{fig:slen}}
\end{figure}
\section{Scattering length}

A crucial parameter affecting low-energy scattering properties
and the positions of high-lying excited states is the s-wave
scattering length $a$. Calculations by Jamieson \emph{et al.}\
\cite{Jamieson:2003} produced values for $a$ ranging from 380
to 1 $a_0$ for the X $^1\Sigma^+$ state of RbCs, depending on
the subtleties of the potential parameters. There are further
uncertainties due to the choice of hyperfine states involved.
In addition, even if the potential is known accurately, there
is the possibility of {\it controlling} the scattering length
with applied fields. We therefore investigate the dependence of
photoassociation on the scattering lengths for both the ground
and excited states.


Scattering lengths were calculated by solving the
Schr{\"o}dinger equation for zero-energy s-wave scattering
numerically using Numerov integration. The long-range
wavefunction was matched to Bessel and Neumann functions of
fractional order \cite{Gribakin:1993}  to take account of the
long-range $R^{-n}$ potential and avoid the need to propagate
to excessively large distances. We adjusted the scattering
length by scaling the whole potential curve by a constant
factor. As is well known, the scattering length is extremely
sensitive to such scaling and passes through a pole whenever
there is a bound state at exactly zero energy. For
$^{85}$Rb$^{133}$Cs a complete cycle is achieved within a
scaling of just over 1\%, as can been seen from Fig.\
\ref{fig:slen}. Rb$_2$ shows similar behavior.

\begin{figure}[tbp]
\includegraphics*[width=\linewidth]{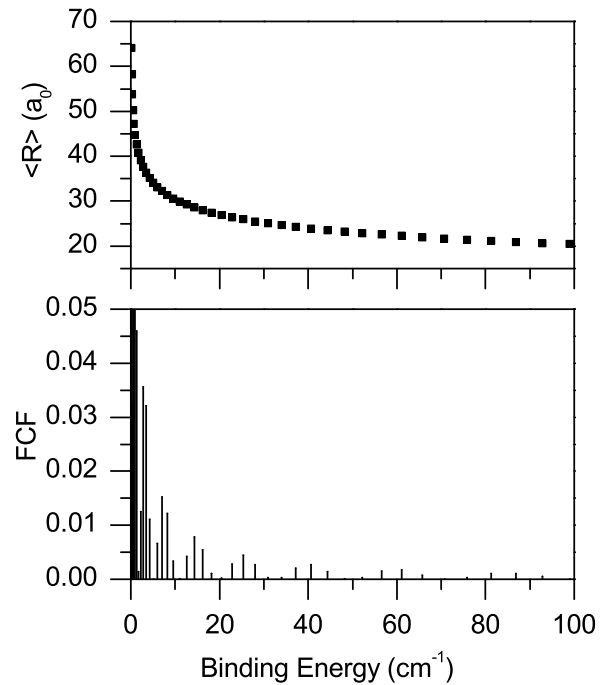}
\caption{Upper panel: Average bond length $\langle R \rangle$
for vibrational levels of RbCs in its A $^1\Sigma^+$ state as a
function of the binding energy. Lower panel: Franck-Condon
factors for photoassociation (in arbitrary units), calculated
for a ground-state scattering length $a=66.3$ $a_0$. Note that
the largest factors are off the top of the scale in this
figure.
\label{fig:FCF:RbCs:general}}
\end{figure}

\begin{figure}[tbp]
\includegraphics*[width=0.95\linewidth]{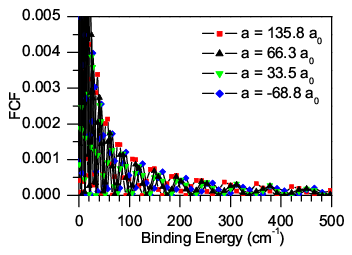}
\includegraphics*[width=0.95\linewidth]{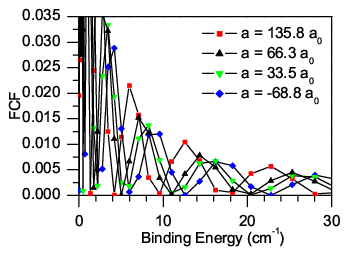}
\includegraphics*[width=0.92\linewidth]{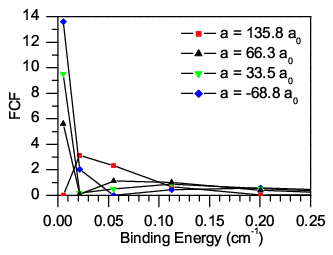}
\caption{Franck-Condon factors for RbCs photoassociation,
calculated for a variety of ground-state scattering lengths.
The scale is arbitrary but is consistent between graphs.
\label{fig:FCF:RbCs:zoom}}
\end{figure}

\section{Franck-Condon factors and their dependence on scattering lengths
\label{sec:FCF}}

We first calculate the Franck-Condon factors (FCFs) of
relevance to photoassociation experiments for
$^{85}$Rb$^{133}$Cs and compare the results with those for the
homonuclear $^{85}$Rb$_2$ system. The purpose of this Section
is to understand the qualitative differences between RbCs and
Rb$_2$, which stem from the different long-range tails of the
excited-state potential (proportional to $R^{-6}$ for RbCs and
$R^{-3}$ for Rb$_2$). These qualitative differences are most
simply illustrated using calculations that neglect the coupling
between the A\,$^1\Sigma^+$ and b\,$^3\Pi$ states, and that is
the approach we use in this Section. However, coupled
calculations of FCFs have been carried out by Luc-Koenig
\cite{Luc-Koenig:private:2008}, and indeed the calculations
described in Section \ref{sec:TD.c} to \ref{sec:TD.e} below use
fully coupled wavefunctions.

The FCFs for photoassociation are calculated as the squares of
overlap integrals between bound vibrational states in the
excited A\,$^1\Sigma^+$ electronic potential and a low-energy
scattering wavefunction on the ground-state potential. Both the
bound and scattering states were obtained by Cooley-Numerov
propagation \cite{Cooley:1961} for rotational angular momentum
$N=0$. A LeRoy-Bernstein analysis \cite{LeRoy:1970} confirmed
that every bound state had been found. The s-wave scattering
function was calculated at a collision energy of 1 mK and
normalized by setting its maximum amplitude to 1 as
$R\rightarrow\infty$. The ground-state potential does not die
off to 1 mK until $R \sim 110\ a_0$, so that the relative
values of the resulting Franck-Condon factors should be valid
for excitation to states dominated by distances $R<100\ a_0$.

Figure \ref{fig:FCF:RbCs:general} shows the calculated FCFs for
RbCs with $a=+66.3$ $a_0$ as a function of binding energy
(detuning from the Cs atomic line), and the corresponding
values of $\langle R \rangle$ for the excited vibrational
states. The FCFs show a typical oscillating pattern as the
bound and continuum states come into and out of phase with one
another. For RbCs each peak of the envelope encompasses around
4 vibrational levels of the excited state. The most intense
peak is just below the dissociation threshold, and the state at
the center of this peak has $\langle R \rangle \approx$ 71
$a_0$. The next peaks in order of decreasing intensity
correspond to $\langle R \rangle \approx$ 47, 37, 32, 29 and 26
$a_0$. The largest FCF occurs for the least-bound state, with
$\langle R \rangle=150$ $a_0$, which is not part of an
envelope. However, this state is bound by only 0.0007
cm$^{-1}$, and it is likely that frequencies this close to the
atomic line would be blocked in an actual experiment.

\begin{figure}[tbp]
\includegraphics*[width=\linewidth]{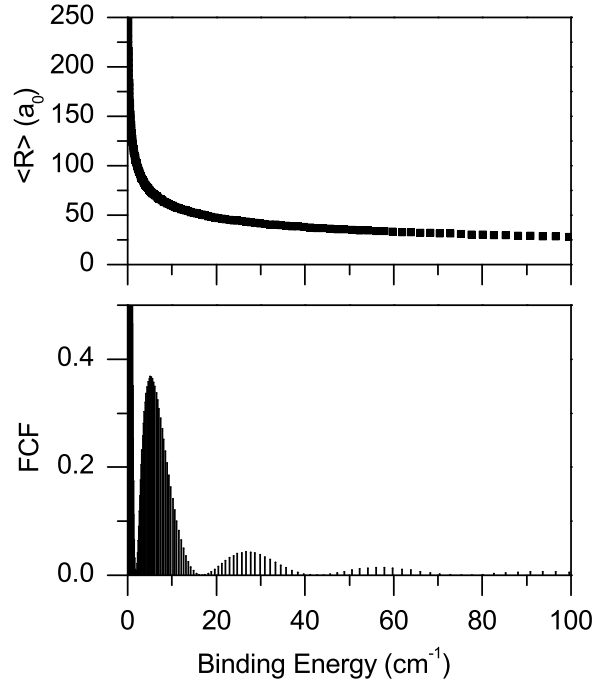}
\caption{Upper panel: Average bond length $\langle R \rangle$
for vibrational levels of Rb$_2$ in its A $^1\Sigma^+_u$ state
as a function of the binding energy. Lower panel: Franck-Condon
factors for photoassociation. The scale is arbitrary but is
consistent with that for RbCs in Figs.\
\ref{fig:FCF:RbCs:general} and \ref{fig:FCF:RbCs:zoom}. Note
that the largest factors are off the top of the scale in this
figure.
\label{fig:FCF:Rb2:general}}
\end{figure}

To investigate the effect of adjusting the ground-state
scattering length, we have repeated the Franck-Condon
calculations for $a= 135.8$, 33.5, and $-68.8$ $a_0$,
corresponding to potential scaling factors of 0.9965, 1.0015
and 1.0039 respectively. The results are shown in Fig.\
\ref{fig:FCF:RbCs:zoom}. It may be seen that changing the
scattering length shifts the positions of the peaks in the
envelope of the intensity distribution. Nevertheless, the peak
intensities themselves follow a curve that is only very weakly
potential-dependent. The peak intensities are thus almost a
single-valued function of binding energy.

\begin{figure}[tbp]
\includegraphics*[width=\linewidth]{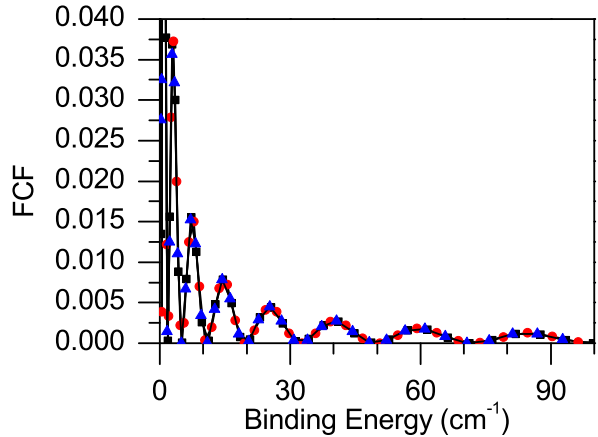}
\caption{Franck-Condon factors for different scalings of the upper
potential with the unscaled ground-state potential. Black, blue
and red symbols are for scalings of 1.0, 1.001 and 1.005
respectively.\label{fig:FCF:a-upper}}
\end{figure}

Scaling the potential curve for the excited state, while
keeping the ground-state scattering length constant, shifts the
positions of the individual {\em lines} but does not alter the
positions of the peaks in the intensity distribution. This
effect is shown for $a=+66.3$ $a_0$ in Fig.\
\ref{fig:FCF:a-upper}: the frequencies that correspond to
vibrational levels shift as the potential is scaled, but the
envelope of the FCFs does not change. Since a laser pulse will
always cover a range of binding energies, it is this envelope
that matters more than the specific position of the levels and
the overall transition probabilities will not be greatly
affected by changes in the excited-state scattering length.

The Franck-Condon factors for photoassociation to form Rb$_2$
are shown in Fig.\ \ref{fig:FCF:Rb2:general}. The potential
curve used here has $a=157.9$ $a_0$. Once again the FCFs show
an oscillatory structure. The major difference from RbCs is
that the oscillations in the envelope of line intensities are
slower as a function of binding energy and the lines themselves
are much more densely packed because of the greater density of
states for an $R^{-3}$ potential. The overall values of the
FCFs are also somewhat larger in the homonuclear case.

\begin{figure}[tb]
\includegraphics*[width=\linewidth]{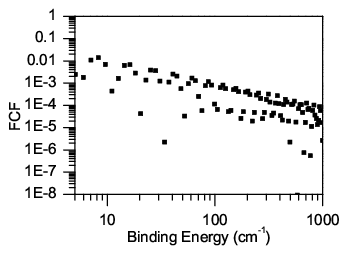}
\includegraphics*[width=\linewidth]{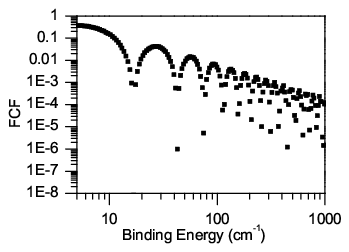}
\caption{Franck-Condon factors for different values of the
ground-state scattering length, shown on a log-log plot for RbCs
(upper panel) and Rb$_2$ (lower panel). Note that the peak values
lie on a straight line in each case, but with a different slope,
corresponding to a different power-law dependence on binding
energy. \label{fig:FCF:loglog}}
\end{figure}

Another significant difference between the heteronuclear and
homonuclear cases is shown in Fig.\ \ref{fig:FCF:loglog}. On a
log-log plot, the intensities of the peak envelopes vary almost
linearly with binding energy $E_{\rm D}-E_v$ near dissociation.
This corresponds to a power-law dependence on binding energy.
However, the powers involved are clearly different for the two
cases: about $(E_{\rm D}-E_v)^{-1}$ for the homonuclear case
and $(E_{\rm D}-E_v)^{-1.6}$ for the heteronuclear case. This
will be significant in designing pulsed-laser experiments,
because the intensity will fall off considerably faster with
detuning (from the atomic line) in the homonuclear case than in
the heteronuclear case. Thus a broad pulse will excite a
wavepacket in which the relative population of deeper levels is
lower in the heteronuclear case.

\section{Time-dependent calculations}

\subsection{Vibrational periods}
In section \ref{sec:waves} we study the photoassociation
process in a time-dependent wavepacket approach. However, we
can make a rough estimate of the time needed for the wavepacket
to evolve from long to short range by considering the
vibrational half-period,
\begin{eqnarray}
t_{1/2} = \frac{1}{2\nu} = \frac{h}{2\left(\frac{dE_v}{dv}\right)},
\end{eqnarray}
where $\nu$ is the classical vibrational frequency. The
vibrational spacing $dE_v/dv$ may be calculated exactly for a
particular potential curve but it is instructive to consider
its value in long-range theory. If only the leading term in the
long-range potential is included, $V(R)=-C_nR^{-n}$, the
vibrational period may be written semiclassically
\cite{LeRoy:1973}
\begin{eqnarray}
\frac{dE_v}{dv} = K_n (E_{\rm D}-E_v)^{(n+2)/2n},
\end{eqnarray}
where
\begin{eqnarray}
K_n = \frac{\overline{K}_n}{\mu^{1/2} C_n^{1/n}}
= \frac{(2\pi)^{1/2}h}{\mu^{1/2} C_n^{1/n}}
\left(\frac{n\Gamma(1+1/n)}{\Gamma(1/2+1/n)}\right),
\end{eqnarray}
$\mu$ is the reduced mass and $\Gamma$ is the gamma function.
Le Roy has tabulated numerical values of the coefficients
$\overline{K}_n$ \cite{LeRoy:1973}. The times obtained for
wavepackets corresponding to each of the main peaks in Fig.\
\ref{fig:FCF:RbCs:general} are given in Table
\ref{tab:FCF:RbCs} for $^{85}$Rb$^{133}$Cs. The half-periods
vary from over 100 ps at small detunings to under 1 ps for a
detuning of $-315$ cm$^{-1}$. Although the positions of the
individual maxima are sensitive to details of the potential,
the time taken is an almost single-valued function of the
binding energy (laser detuning) for a particular long-range
potential and reduced mass.

\begin{table}[tbp]
\caption{Vibrational half-periods for an RbCs molecule at
energies corresponding to maxima in the intensity distribution
for photoassociation. The index $m$ labels the successive
maxima, starting at the dissociation limit.
\label{tab:FCF:RbCs}}
\begin{tabular}{ccccc}
\hline\hline
$m$ & \quad binding energy \quad
& \quad $\langle R \rangle$ \quad & \quad relative \quad & \quad $t_{1/2}$ \\
& (cm$^{-1}$) & ($a_0$) & intensity & (ps) \\ \hline
$1$ & $-0.0063$ & 71 & $\sim1.5$ & 290 \\
$2$ & $-0.76$ & 47 & 0.13 & 54 \\
$3$ & $-2.9$ & 37 & 0.038 & 22 \\
$4$ & $-7.3$ & 32 & 0.016 & 12 \\
$5$ & $-15$ & 29 & 0.0080 & 7.5 \\
$6$ & $-26$ & 26 & 0.0045 & 5.2 \\
$7$ & $-41$ & 24 & 0.0027 & 3.8 \\
$8$ & $-57$ & 23 & 0.0018 & 3.0 \\
$9$ & $-82$ & 20 & 0.0013 & 2.4 \\
$10$ & $-113$ & 20 & 0.00096 & 1.93 \\
$11$ & $-150$ & 19 & 0.00073 & 1.56 \\
$12$ & $-194$ & 18 & 0.00056 & 1.26\\
$13$ & $-245$ & 17 & 0.00041 & 1.13 \\
$14$ & $-315$ & 17 & 0.00028 & 0.97 \\
\hline\hline
\end{tabular}
\end{table}

The times taken for RbCs are significantly shorter than those
for Rb$_2$, shown in Table \ref{tab:FCF:Rb2}. For example, for
a detuning of 25 cm$^{-1}$ from the atomic line, the classical
time is about 13 ps for Rb$_2$ but 5 ps for RbCs.

The classical model used here is of course considerably
oversimplified and neglects features such as the spreading of a
wavepacket as it propagates inwards. A full treatment of the
time evolution requires quantum-mechanical calculations as
described below.

\begin{table}[tbp]
\caption{Vibrational half-periods for an Rb$_2$ molecule at
energies corresponding to maxima in the intensity distribution
for photoassociation. The index $m$ labels the successive
maxima, starting at the dissociation limit.
\label{tab:FCF:Rb2}}
\begin{tabular}{ccccc}
\hline\hline
$m$ & \quad binding energy \quad
& \quad $\langle R \rangle$ \quad & \quad relative \quad & \quad $t_{1/2}$ \\
& (cm$^{-1}$) & ($a_0$) & \quad intensity \quad & \quad (ps) \quad \\ \hline
$1$ & $-0.017$ & 500 & 12 &  6300 \\
$2$ & $-0.038$ & 380 & 8.5 & 3200 \\
$3$ & $-0.10$ & 270 & 5.2 & 1400 \\
$4$ & $-0.49$ & 160 & 2.4 & 380 \\
$5$ &  $-5.1$ & 75 & 0.36 & 54 \\
$6$ & $-27$ & 43 & 0.043 & 13 \\
$7$ & $-56$ & 34 & 0.014 & 7.3 \\
$8$ & $-91$ & 29 & 0.007 & 4.9 \\
$9$ & $-135$ & 26 & 0.004 & 3.5 \\
\hline\hline
\end{tabular}
\end{table}

\subsection{Wavepacket model and methods\label{sec:waves}}

In the time-dependent wavepacket approach, the photoassociation
reaction starts with two cold atoms in the ground electronic
state with relative kinetic energy $E_{\rm kin}$. Absorption of
a photon red-detuned by an energy $\hbar\omega_{L}$ from the
atomic resonance line at $\hbar\omega_{\rm at}$ produces a
molecule in the excited electronic state as shown in Fig.\
\ref{scheme}. The Hamiltonian describing optical transitions
between the X\,$^1\Sigma^+$ electronic ground state and the
coupled A\,$^1\Sigma^+$ and b\,$^3\Pi$ excited states can be
represented in the diabatic basis as
\begin{widetext}
\begin{eqnarray}
\Op{H} = \left( \begin{array}{ccc}
\Op{T}
+ V_{{\rm X}^1\Sigma^+}(R) & \bm{\mu}(R)\cdot\bm{E}(t) & 0 \\
\bm{\mu}(R)\cdot\bm{E}(t) & \Op{T}
+ V_{{\rm A}^1\Sigma^+}(R) -
\hbar\omega_L & \sqrt{2}W^{\Sigma\Pi}_{\rm SO}(R) \\
0 & \sqrt{2}W^{\Sigma\Pi}_{\rm SO}(R) & \Op{T}
+ V_{{\rm b}^3\Pi}(R)
- W^{\Pi\Pi}_{\rm SO}(R) -\hbar\omega_L \end{array} \right)\,,
\label{hamil}
\end{eqnarray}
\end{widetext}
where $\Op{T}$ is the kinetic energy operator and $V_i(R)$ are
the respective potential energy curves. Assuming the dipole and
rotating-wave approximations, the coupling between the
X\,$^1\Sigma^+$ and A\,$^1\Sigma^+$ electronic states is
represented by the scalar product between the transition dipole
moment, $\bm{\mu}(R)$, and the polarization vector of the laser
field, $\bm{E}(t)$. In the present work the $R$-dependence of
$\bm{\mu}(R)$ is neglected and it is represented by its
asymptotic value. The A\,$^1\Sigma^+$ and b\,$^3\Pi$ excited
states are coupled by the spin-orbit interaction. The hyperfine
interaction can be neglected because the associated timescale
is much longer than the femtosecond or picosecond laser pulses,
so that the hyperfine interaction is not resolved for the
processes considered in the present study.

The Hamiltonian, Eq.~(\ref{hamil}), is represented on a Fourier
grid with a variable grid step \cite{kokoouline:JCP:1999}.
Details of the mapped Fourier Grid Hamiltonian (FGH) method are
described in refs.\ \onlinecite{Willner:JCP:2004} and
\onlinecite{Kallush:CPL:2006}. Using the FGH method, we are
able to extend the spatial grid to $R \sim 10^3\ a_0$ using
only 1023 grid points; this is sufficient both to represent the
bound vibronic levels and to approximate the scattering
continuum in the ultracold temperature regime by box states
\cite{Koch:JPhysB:2006}.

To describe the short-pulse photoassociation process, we solve
the time-dependent Schr\"odinger equation (TDSE),
\begin{equation}
i\hbar\frac{\partial}{\partial t}  \Psi(t) =
\Op{H}(t)\Psi(t),
\end{equation}
by expanding the evolution operator $\exp[-i\Op{H}t/\hbar]$ in
terms of Chebychev polynomials
\cite{kosloff:AnnRevPhysChem:94}. Because of the large extent
of the grid and the very small kinetic energy of the system, it
is not necessary to enforce an absorbing boundary condition at
the edge of the grid.

\subsection{Analysis of the coupled vibronic wavefunctions\label{sec:TD.c}}

Diagonalization of the Hamiltonian of Eq.~(\ref{hamil}) with
$\bm{E}(t)$ set to zero yields the binding energies and
vibronic wavefunctions and allows the calculation of FCFs,
rotational constants, etc. For RbCs it produces 136 bound
vibronic levels for the X\,$^1\Sigma^1$ ground electronic state
and 398 levels for the coupled A\,$^1\Sigma^+$ and b\,$^3\Pi$
electronic states. The eigenfunctions of the excited electronic
states are perturbed by the spin-orbit coupling and have mixed
singlet and triplet character. The highest excited-state level
has a binding energy of 0.002$\,$cm$^{-1}$, which differs
slightly from the result in Section \ref{sec:FCF} because the
latter neglected the coupling between the singlet and triplet
states.

When two electronic states interact and their curves cross,
there are two quite different limiting cases that may be
considered to be ``uncoupled''. If the coupling near the
crossing point is very weak (compared to the local vibrational
spacings), the crossing is only weakly avoided and the
eigenstates of the coupled system are close to those of the
individual diabatic (crossing) electronic states. Conversely,
if the coupling is strong, the crossing is strongly avoided and
the best zeroth-order picture is to consider eigenfunctions of
the {\em adiabatic} states, defined by the non-crossing upper
and lower adiabatic curves. In this case the adiabatic states
themselves change character from one side of the crossing to
the other, so that in the present case the zeroth-order states
are predominantly singlet on one side of the crossing and
predominantly triplet on the other. The couplings between the
adiabatic states are provided by nonadiabatic couplings (which
are related to off-diagonal matrix elements of $d/dR$ and
$d^2/dR^2$).

\begin{figure*}[tbp]
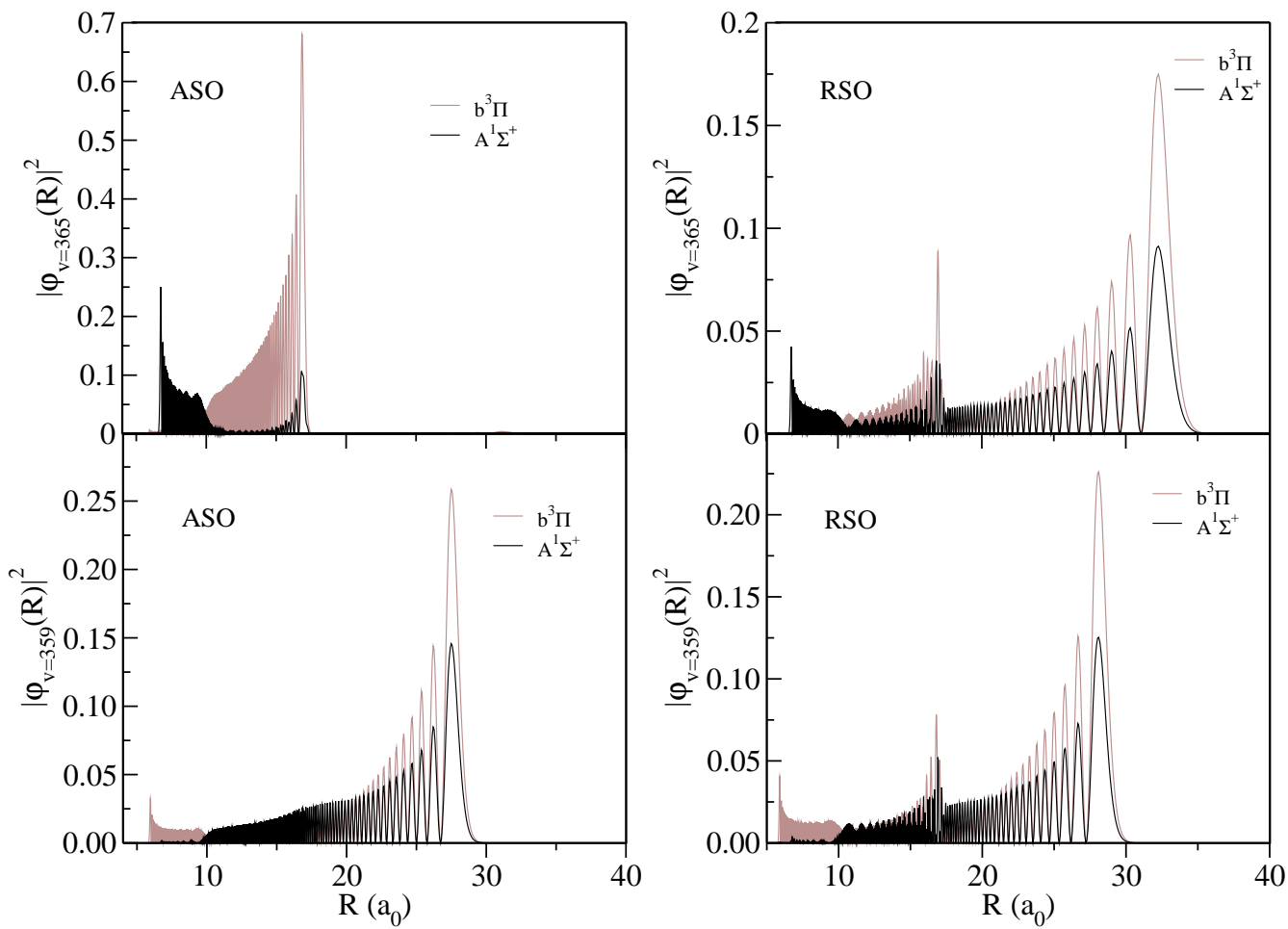

\subfigure{\includegraphics[width=0.49\linewidth]{figures/eigen_ASO}}
\subfigure{\includegraphics[width=0.49\linewidth]{figures/eigen_RSO}}
\caption{Eigenfunctions of the coupled excited-state potentials
corresponding to $v=359$ and 365. Left-hand panels: results
obtained with constant asymptotic spin-orbit coupling (ASO).
Right-hand panels: results obtained with $R$-dependent
spin-orbit coupling (RSO). \label{fig:eigen}}
\end{figure*}

The wavefunction for a mixed vibronic state may be written
\begin{equation} \varphi_1^v(R)\psi_1 +
\varphi_3^v(R)\psi_3 \label{eq:mixed}
\end{equation}
where the functions $\varphi_1^v(R)$ and $\varphi_3^v(R)$ are
vibrational wavefunctions and $\psi_1$ and $\psi_3$ are
electronic functions for the singlet and triplet excited states
respectively. It is useful to compare the eigenstates that are
obtained for two different cases, as shown in Fig.\
\ref{fig:eigen}. The two panels on the left show RbCs
eigenstates calculated with the spin-orbit coupling functions
fixed at their asymptotic value, $\Delta E_{\rm SO}^{\rm Cs} =
554.04$ cm$^{-1}$ (corresponding to $W_{\rm
SO}^{\Sigma\Pi}=184.68$ cm$^{-1}$, which is considerably larger
than the experimentally derived value \cite{Fellows:priv:2007}
around the crossing point of the diabatic curves ($W_{\rm
SO}^{\Sigma\Pi}\approx 100$ cm$^{-1}$, see Fig.\ \ref{pes}).
These will be referred to as ASO (asymptotic spin-orbit)
calculations. The two levels shown are for vibrational indices
$v'=359$, with binding energy 22.19 cm$^{-1}$, and $v'=365$,
with binding energy 10.01 cm$^{-1}$. It may be seen that for
both ASO wavefunctions there is a switchover from singlet to
triplet character around the avoided crossing ($R \sim 10\
a_0$), but in opposite directions. This occurs because the
states are nearly adiabatic: the function for $v'=365$ lies on
the upper adiabatic curve, so has mainly singlet character
inside the crossing and mainly triplet character outside. The
function for $v'=359$ lies on the lower adiabatic curve, so
that it has mostly triplet character at short range. Outside
the crossing, it has mainly singlet character in the region
where the singlet and triplet curves are separated by more than
the spin-orbit coupling. At long range ($R> {\sim 20}\ a_0$),
however, it reacquires the triplet admixture characteristic of
an $\Omega=0$ state at the 5S + 6P$_{1/2}$ threshold, yielding
a 2:1 ratio of triplet to singlet probability. It stretches to
considerably greater internuclear distance because of the
different turning points for the two adiabatic states.

The near-adiabatic vibronic levels obtained for the ASO case
provide a context to understand the eigenstates obtained with
the $R$-dependent spin-orbit (RSO) coupling function, shown in
the panels on the right of Fig.\ \ref{fig:eigen}. In the RSO
case the levels with index numbers $v'=359$ and 365 have
binding energies 17.79 cm$^{-1}$ and 7.53 cm$^{-1}$,
respectively. The $v'=359$ level exists {\em mostly} on the
lower adiabatic curve, and has triplet character with only a
small singlet admixture at short range and switches over to
singlet character (with the same characteristic triplet
admixture as before) outside $R=10\ a_0$. Nevertheless, there
is significant nonadiabatic mixing which introduces upper-state
character, shown by the peaks in probability densities around
the outer turning point of the upper curve near $R \sim17\
a_0$. The $v'=365$ level is more strongly mixed, with a larger
contribution from the upper adiabatic state that makes it
predominantly singlet at short range. However, this state too
has substantial lower-state character so that it does have
density out to the outermost turning point ($R \sim33\ a_0$ in
this case). It is notable that {\em both} RSO eigenstates have
significantly enhanced singlet density near the outer turning
point of the upper curve ($R \sim17\ a_0)$. This is the feature
that in Rb$_2$ was responsible for enhanced deexcitation to
levels of the ground state bound by up to 10 cm$^{-1}$
\cite{Koch:PRA:2006b, Pechkis:PRA:2007}. However, the
wavefunction for $v'=365$ also has substantial singlet density
near the inner turning point of the singlet state; this is a
feature that was {\em not} observed for Rb$_2$ (see Fig.\ 2 of
ref.\ \onlinecite{Koch:PRA:2006b}), and as will be seen below
it allows deexcitation to even more deeply bound levels. The
fact that the wavefunction can have significant singlet density
{\em both} at this turning point {\em and} at the outermost
turning point arises from nonadiabatic coupling between the
upper and lower adiabatic states.

In an alternative approach, the nonadiabatic mixings could be
shown in an adiabatic representation, displaying the components
of the vibrational wavefunctions on the two $0_u^+$ states.
This would emphasize the nearly adiabatic character of the ASO
wavefunctions but would hide their switchover from singlet to
triplet character near the avoided crossing.

\begin{figure}[tbp]
\includegraphics*[width=0.95\linewidth]{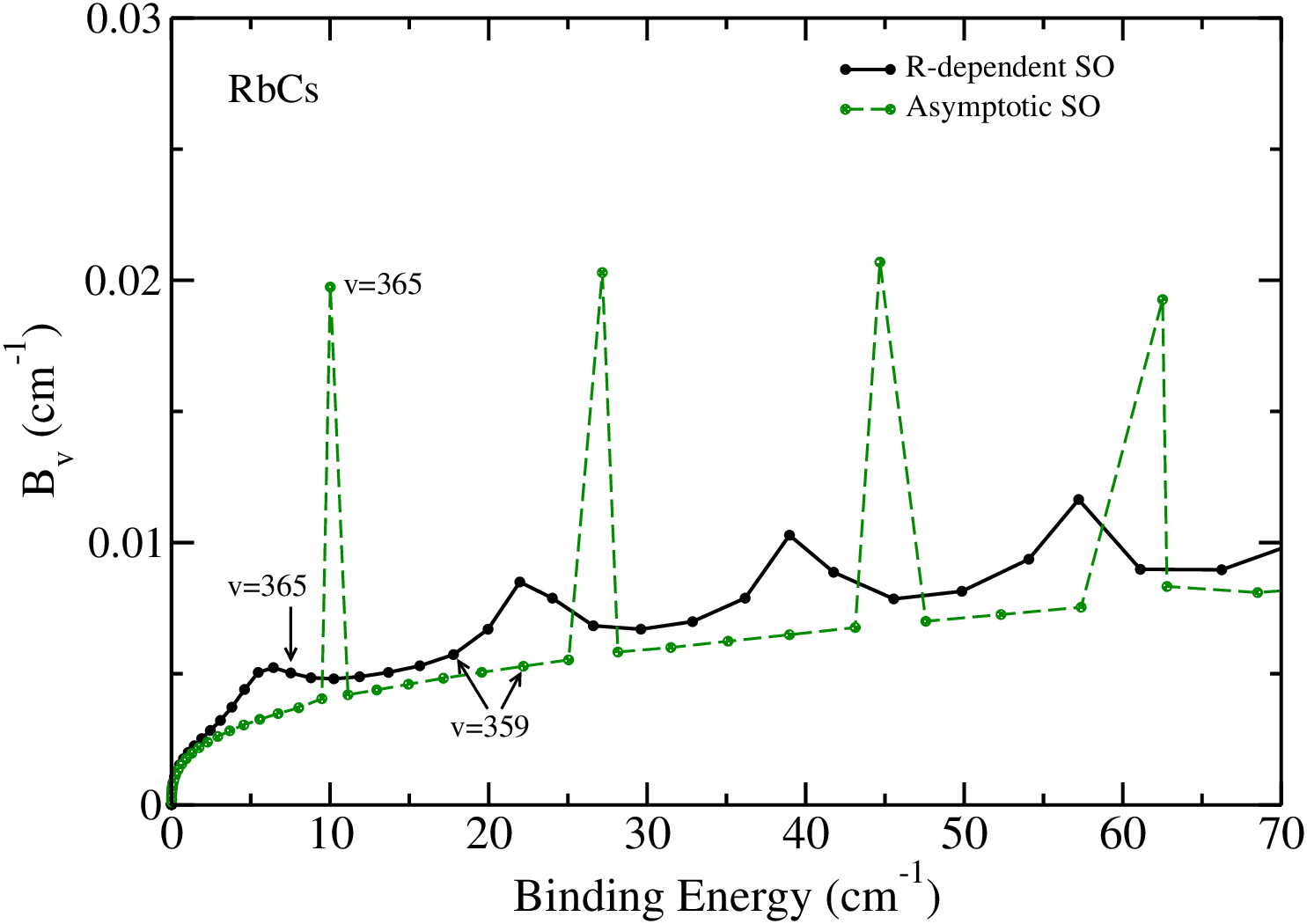}
\caption{Rotational constants $B_v' = \langle h/(8\pi^2c\mu R^2)
\rangle$ corresponding to the vibronic levels of the
coupled A\,$^1\Sigma^+ + b^3\Pi$ excited electronic states for
RbCs.
\label{rotconst}}
\end{figure}

The character of the mixed levels, as a function of vibrational
quantum number, may be seen very clearly in the calculated
rotational constants for the coupled levels ($B_v = \langle
h/(8\pi^2c\mu R^2)\rangle$), which are shown in Fig.\
\ref{rotconst} for both the ASO and RSO cases. In the ASO case,
the rotational constants form two almost independent series,
with the higher values corresponding to levels that are
predominantly on the upper adiabatic curve. The upper state has
a larger vibrational spacing than the lower state. In the RSO
case, the peaks in rotational constants correspond to levels
with a significant {\em contribution} from the upper adiabatic
state. However, it may be seen that in this case the
nonadiabatic coupling is strong enough to spread the character
of each upper-curve vibronic state across several
eigenfunctions of the coupled problem, so that the peaks are
not very large. There are no eigenstates of the coupled problem
that are {\em predominantly} on the upper adiabatic curve.

\begin{figure}[tbp]
\includegraphics*[width=0.9\linewidth]{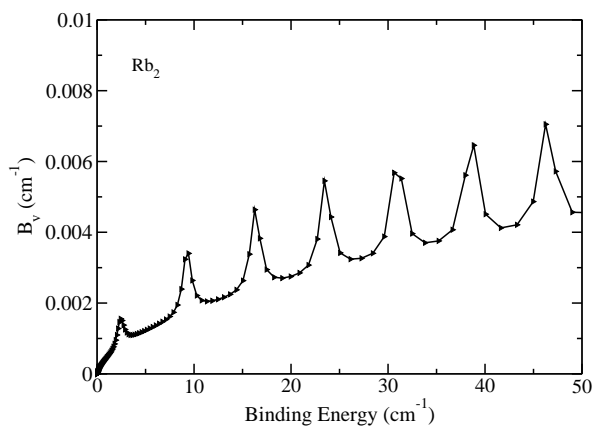}
\caption{Rotational constants $B_v' = \langle h/(8\pi^2 c\mu
R^2) \rangle$ corresponding to the vibronic levels of the
coupled A\,$^1\Sigma^+ + b^3\Pi$ excited electronic states of
Rb$_2$. \label{rotconst-rb2}}
\end{figure}

The rotational constants for RbCs may be compared with those
for Rb$_2$ \cite{kokoouline:JCP:1999,Bergeman:2006}, shown in
Fig.\ \ref{rotconst-rb2}. Rb$_2$ exhibits nonadiabatic mixing
that is intermediate between the ASO and RSO cases for RbCs.
For Rb$_2$ the extent of the mixing has been shown to be
isotope-dependent \cite{Fioretti:JPB:2007}.

The existence of bound states with singlet character both at
long range and near the inner turning point will also be
important in designing fixed-frequency schemes to populate
deeply-bound ground-state levels by STIRAP and related methods.
Such schemes will be most efficient when they proceed via mixed
levels with population on both the upper and lower adiabatic
curves.

\subsection{Photoassociation with a short laser pulse}
\begin{figure}[tbp]
\includegraphics*[width=0.95\linewidth]{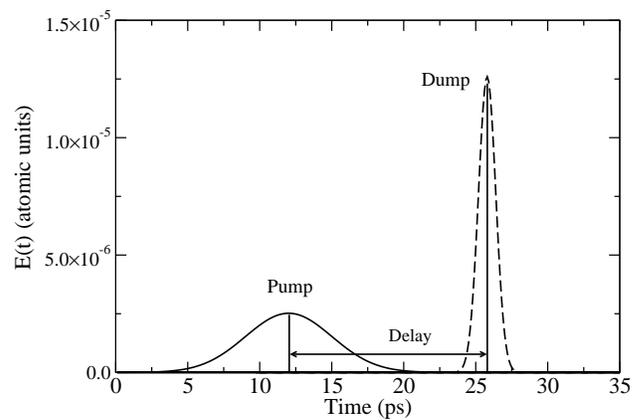}
\caption{Schematic diagram of the laser pulse sequence
considered in the present study. The pump and dump pulses are
centered at 12.0 ps and 25.8 ps, respectively, with a time
delay of 13.8 ps (note that $1\times 10^{-5}$ atomic units of
electric field correspond to 5.14 MV/m).\label{pulse}}
\end{figure}

The pump-dump scheme considered here is shown schematically in
Fig.\ \ref{pulse}. Each laser pulse is assumed to be
transform-limited,
\begin{equation}
E(t) = E_0 f(t) \cos(\omega_Lt),
\end{equation}
with a Gaussian profile $f(t)=\exp-\alpha(t-t{\rm c})^2$. The
pulse has central frequency $\omega_L/(2\pi)$ and maximum field
amplitude $E_0$. The full width at half maximum (FWHM) of the
intensity profile $E_0^2 f(t)^2$ is $\tau_L=(2\ln
2/\alpha)^{-1/2}$. The corresponding FWHM frequency width is
$\delta_\omega/(2\pi)=2\ln 2/(\pi\tau_L)$. The Gaussian
envelopes are centered at times $t=t_{\rm p}$ and $t_{\rm d}$
for the pump and dump pulses respectively.

\begin{table}[tbp]
\caption{The parameters for the pump and dump pulse: detuning
$\delta_L^{\rm at}$ from the atomic line, intensity $I_L$,
temporal width $\tau_L$, spectral bandwidth
$\delta_\omega$, and integrated pulse energy per area.}
\label{tab:pulse}
\begin{tabular}{cccccc}
\hline\hline
pulse & $\delta_L^{\rm at}$ & $I_L$ & $\tau_L$ & $\delta_\omega/(2\pi c)$ & Energy/Area \\
&(cm$^{-1}$)&(W cm$^{-2}$) &(ps)&(cm$^{-1}$)&(J m$^{-2}$)\\
\hline
pump (ASO) &$-10.01$ & 16.86 & 5.00 & 2.94 & 0.095  \\
pump (RSO) &$-7.53$ & 16.86 & 5.00 & 2.94 & 0.095  \\
dump (RSO) & 1427.61 & 98.32 & 1.00 & 14.72 & 0.474 \\
\hline\hline
\end{tabular}
\end{table}

The parameters used for the laser pulses in this study are
given in Table \ref{tab:pulse}. We choose the parameters of the
pump pulse to excite a few excited-state vibronic levels
$v^\prime$ close to the maxima in Fig.\ \ref{rotconst} where
nonadiabatic coupling is strongest. The central frequency is
chosen to be resonant with the level $v^\prime= 365$, with a
laser detuning from the atomic line $\delta_L^{\rm at} = 7.53$
cm$^{-1}$ in the RSO case and 10.01 cm$^{-1}$ in the ASO case.
The temporal width of the pulse is chosen to give an energy
spread that will excite 5 to 10 levels near $v^\prime = 365$.
The intensity is chosen to obtain maximum excitation of the
resonant level.

The initial state $\Psi(t=0)$ is chosen to be a box-quantised
eigenfunction of the field-free Hamiltonian that represents an
s-wave scattering state of the ground-state potential curve
with energy corresponding to a temperature $T=16.5\ \mu$K.

The wavepacket at time $t$ may be written
\begin{equation}
\Psi(t) = \sum_i \Phi_i(R,t)\psi_i\quad i={\rm g},1,3,
\end{equation}
where g indicates the ground electronic state and 1 and 3 refer
to the singlet and triplet excited states. The wavepacket
dynamics are analyzed by studying the evolution of the
population on the respective states,
\begin{equation}
  \label{eq:pop}
  P_i(t) = |\langle\Phi_i(R,t)\vert\Phi_i(R,t)\rangle|^2.
\end{equation}
We also project the wavepacket onto the vibronic wavefunctions
of the ground and coupled excited states,
\begin{eqnarray}
  \label{eq:td-FCF}
  P_{\rm g}^{v''}(t) &=& |\langle \varphi_{\rm g}^{v''}(R)|\Phi_1(R,t)\rangle|^2 \\
  \label{eq:proj}
  P_{\rm e}^{v'}(t) &=& \Bigl| \sum_{i=1,3} \langle \varphi_i^{v'}(R)|\Phi_i(R,t)\rangle\Bigr|^2.
\end{eqnarray}
where the vibronic function on the coupled excited states is
given by Eq.\ (\ref{eq:mixed}). Since the amplitude of
$\Psi(t=0)$ at short range depends on the size of the box used
(extent of the FGH grid, $R_{\rm max}$), the absolute values of
the populations and projections decrease approximately linearly
with $R_{\rm max}$.

\begin{figure}[tbp]
\includegraphics*[width=0.95\linewidth]{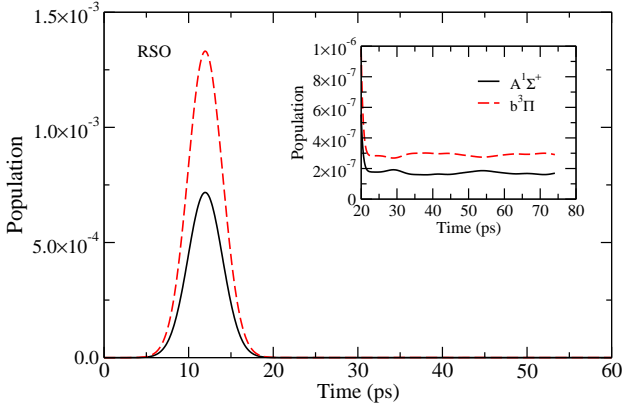}

\caption{The total populations $P_i(t)$ of the excited
A\,$^1\Sigma^+$ and b\,$^3\Pi$ electronic states during the RSO
pump pulse, relative to the ground-state population. The
fractions of the population that remain in the excited states
after the pulse are shown in the inset. \label{norm}}
\end{figure}
The overall populations of the two electronically excited
states, $P_i(t)$, are shown in Fig.\ \ref{norm} for the RSO
case. They reach a maximum near $\tau_{\rm p}$ = 12.0 ps.
Before the end of the pump pulse, most of the population
returns to the initial continuum state. This corresponds to
coherent transients \cite{Allen:1974} which are off-resonant
and can be excited only during the pulse. Only a small amount
($\sim 5\times 10^{-7}$) remains in the vibronic levels of the
coupled excited states (as shown in the inset of Fig.\
\ref{norm}).
This population continues to oscillate between the two
electronically excited states due to the nonadiabatic coupling.
Such oscillations have also been observed for Rb$_2$
\cite{Koch:PRA:2006b,Mur-Petit:PRA:2007}.
The results for the ASO case are qualitatively
similar, but reduced by about a factor of 2 because of the
larger detuning from the atomic line.

\begin{figure*}[tbp]
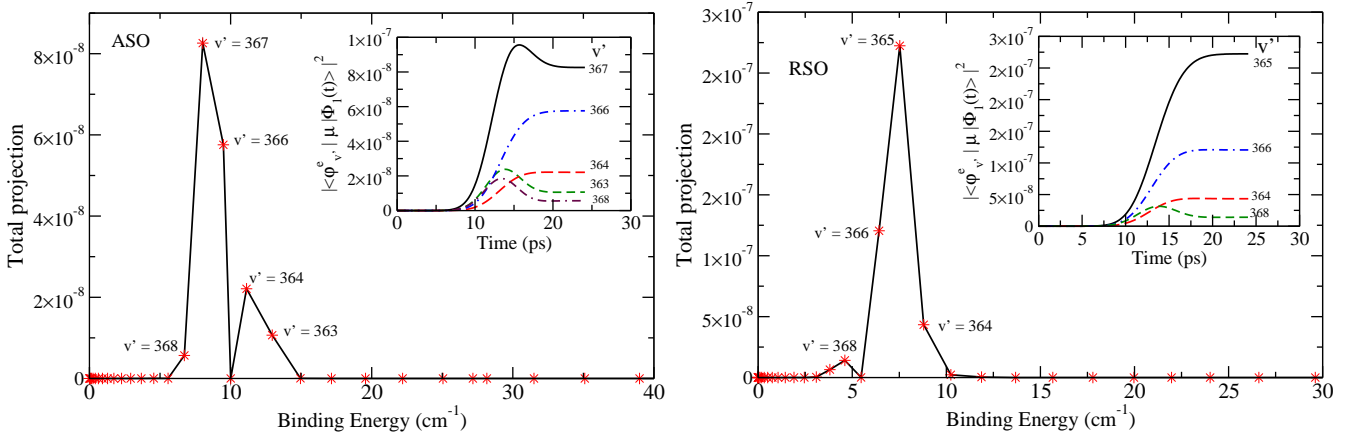

\subfigure{\includegraphics*[width=0.49\linewidth]{figures/proje2e_ASO}}
\subfigure{\includegraphics*[width=0.49\linewidth]{figures/proje2e_RSO}}
\caption{The final projections $P_{\rm e}^{v'}(t)$ of the
wavepacket onto the excited-state vibronic eigenfunctions as a
function of their binding energies. Significant populations of
molecules are formed in vibronic levels $v^\prime = 363$ to
370. The time-dependence during the pulse is shown in the
insets for the levels with the largest final population.
Left-hand panel: results obtained with asymptotic spin-orbit
coupling (ASO) Right-hand panel: results obtained with
$R$-dependent spin-orbit coupling (RSO). \label{proje2e}}
\end{figure*}

Detailed information about the population of the individual
vibronic levels $v'$ can be obtained from the projection of the
excited wavepacket onto excited-state vibronic levels
$P^{v'}_{\rm e}(t)$. Many vibronic levels are excited
transiently during the pulse, but only levels with $v^\prime =
363$ to 370 remain significantly populated after the pulse. The
final populations of the individual near-resonant vibronic
levels are shown in Fig.\ \ref{proje2e}, with the
time-dependence of the populations during the pulse shown in
the insets. The final populations peak near $v^\prime$ = 365 in
both cases, but with variations arising from differences in
FCFs and for levels that are further off-resonance. It is
noteworthy that $v'=365$ itself is almost unpopulated in the
ASO case: it has a very small FCF because it resides on the
upper adiabatic curve and has very little probability density
near the outermost turning point. In the RSO case, all the
levels have significant upper-state character: in this case the
anomalously low population of $v^\prime$ = 367 occurs because
the maximum of the last peak of the eigenfunction corresponds
to a node in the initial ground-state wavefunction at $R \sim
34.7\ a_0$.

In order to determine how many molecules are formed per
photoassociation pulse, it is necessary to average over the
final excited-state populations obtained from all thermally
populated initial scattering states \cite{Koch:absolute:2006}.
The averaging procedure relates the scattering states employed
in the calculations to the actual volume of the trap. The
absolute number of molecules is limited by the probability
density of atoms pairs near the Condon radius, i.e. by the
population within the `photoassociation window'. While the
exact value depends on the details of the potential, and in
particular on the scattering length, an order-of-magnitude
estimate can be obtained by comparing simulations for RbCs and
Rb$_2$ using a similar scattering length, the same size of the
grid and an identical initial state. In order to compare to the
results for Rb$_2$ \cite{Koch:PRA:2006b, Koch:absolute:2006},
we have repeated the simulation for RbCs shown in Fig.\
\ref{norm} using a grid of 20000 a$_0$ and an initial state
with scattering energy corresponding to 100 $\mu$K. The final
excited-state norm of $1.1\times 10^{-7}$ is about two orders
of magnitude smaller than for Rb$_2$ (cf. Table I of Ref.\
\onlinecite{Koch:PRA:2006b}), where about one molecule per
pulse is predicted for $10^{8}$ atoms at a density of
$10^{-10}$ cm$^{-3}$ \cite{Koch:absolute:2006}. The reduced
photoassociation yield of a $1/R^6$ potential compared to a
$1/R^3$ potential requires a significantly higher density
and/or a larger number of atoms to produce a comparable
population of ground-state molecules.

\subsection{Formation of ultracold ground-state RbCs molecules\label{sec:TD.e}}

Some very loosely bound ground-state vibrational levels are
populated directly by the pump pulse in processes involving
more than one photon, but a dump pulse is required to populate
more deeply bound levels. The dump pulse is activated after an
appropriate time delay, when a substantial amount of the
wavepacket has reached the short-range region where there is
significant overlap with deeply bound levels of the ground
electronic state. Since coherent effects between the pump and
dump pulses can be neglected, the pump and dump pulses are
treated separately. The excited-state population is normalized
to one after the pump pulse in order to have a direct
correspondence between the populations calculated and the
probability of formation of ground state molecules.

\begin{figure*}[tbp]
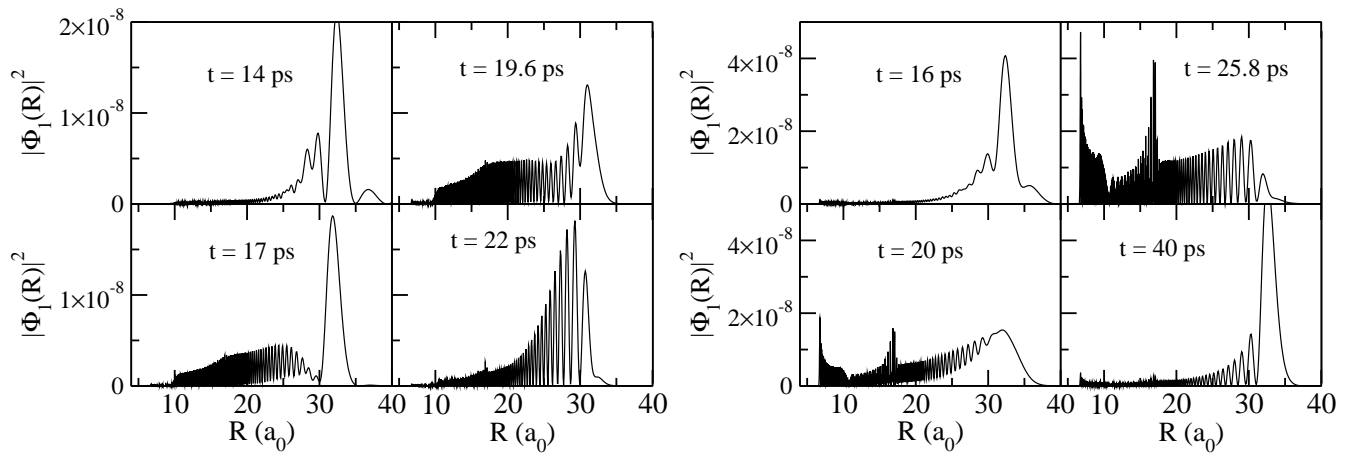

\subfigure{\includegraphics*[width=0.49\linewidth]{figures/snap_ASO}}
\subfigure{\includegraphics*[width=0.49\linewidth]{figures/snap_RSO}}
\caption{Snapshots of the A\,$^1\Sigma^+$ component of the
wavepackets at different times after the pump pulse. Left-hand
panels: results obtained with asymptotic spin-orbit coupling
(ASO). Right-hand panels: results obtained with $R$-dependent
spin-orbit coupling (RSO).} \label{snap}
\end{figure*}

It is useful to inspect the probability density distribution of
the excited-state wavepackets as a function of time. Snapshots
of the A\,$^1\Sigma^+$ component of the wavepackets for a few
representative times are shown in Fig.\ \ref{snap} for both the
ASO and RSO cases. The wavepacket is initially quite similar in
the two cases and starts to move inwards under the influence of
the long-range potential, slightly faster in the ASO case
because of the larger detuning from the atomic line. However, a
major difference arises after the wavepackets reach the
crossing point between the diabatic curves. The ASO wavepacket
remains almost entirely on the lower adiabatic curve, and thus
has mostly triplet character inside $10\,a_0$. By contrast, the
RSO wavepacket is made up of levels of mixed upper and
lower-state character and there is a large buildup of singlet
probability density near the inner turning point. This reaches
a maximum at about 26 ps. Movie of the propagation of the
A\,$^1\Sigma^+$ and b\,$^3\Pi$ components of the wavepackets
are.

\begin{figure}[tbp]
\caption{Movie showing the A\,$^1\Sigma^+$ and b\,$^3\Pi$
component of the wavepackets as a function of time for the
asymptotic spin-orbit coupling (ASO) coupling function.
[Animated gif file movieASO.gif]} \label{movie_ASO}
\end{figure}

\begin{figure}[tbp]
\caption{Movie showing the A\,$^1\Sigma^+$ and b\,$^3\Pi$
component of the wavepackets as a function of time for the
$R$-dependent spin-orbit coupling (RSO) coupling function.
[Animated gif file movieRSO.gif]} \label{movie_RSO}
\end{figure}

\begin{figure*}[tbp]
\subfigure{\includegraphics*[width=0.49\linewidth]{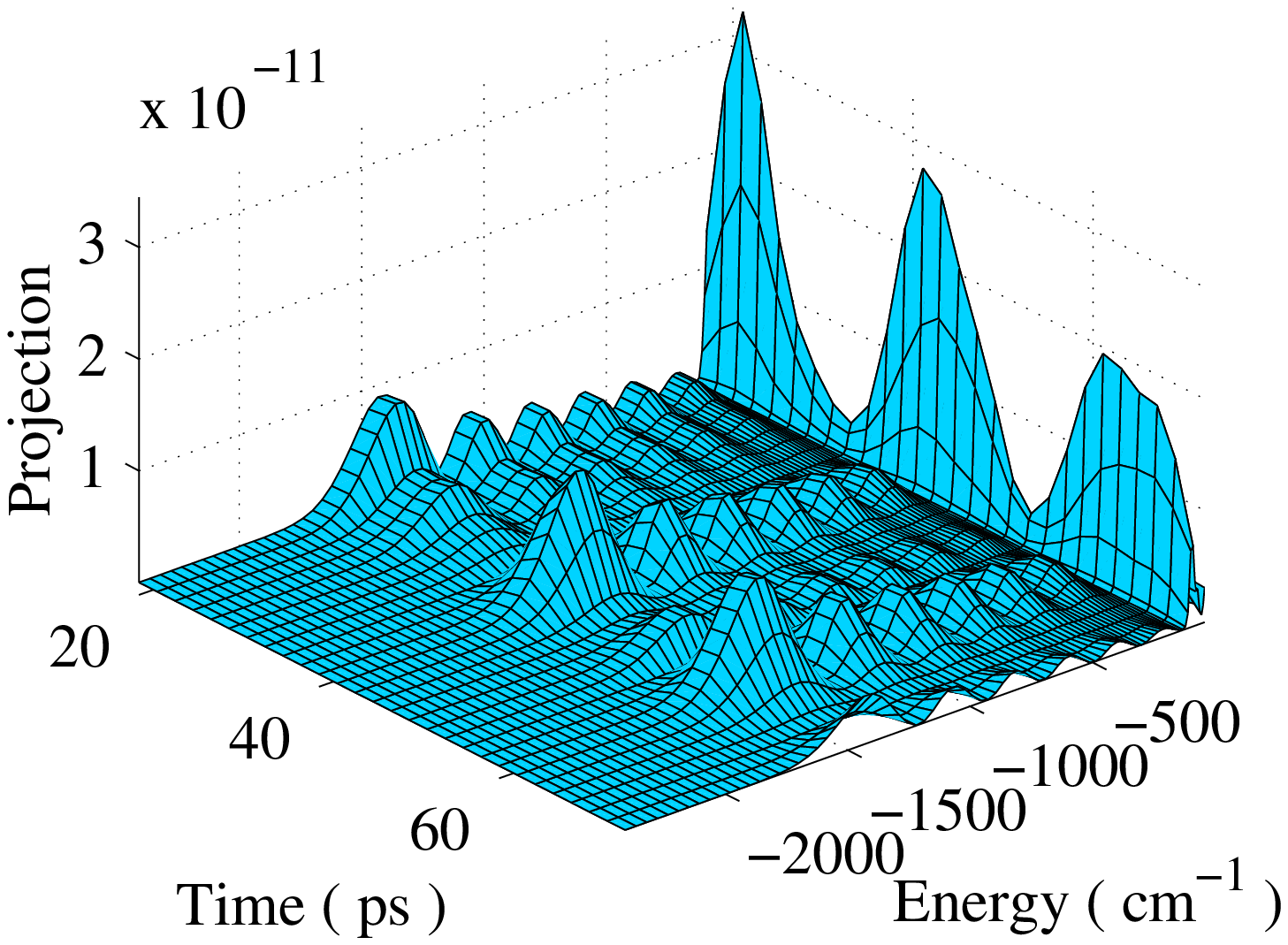}}
\subfigure{\includegraphics*[width=0.49\linewidth]{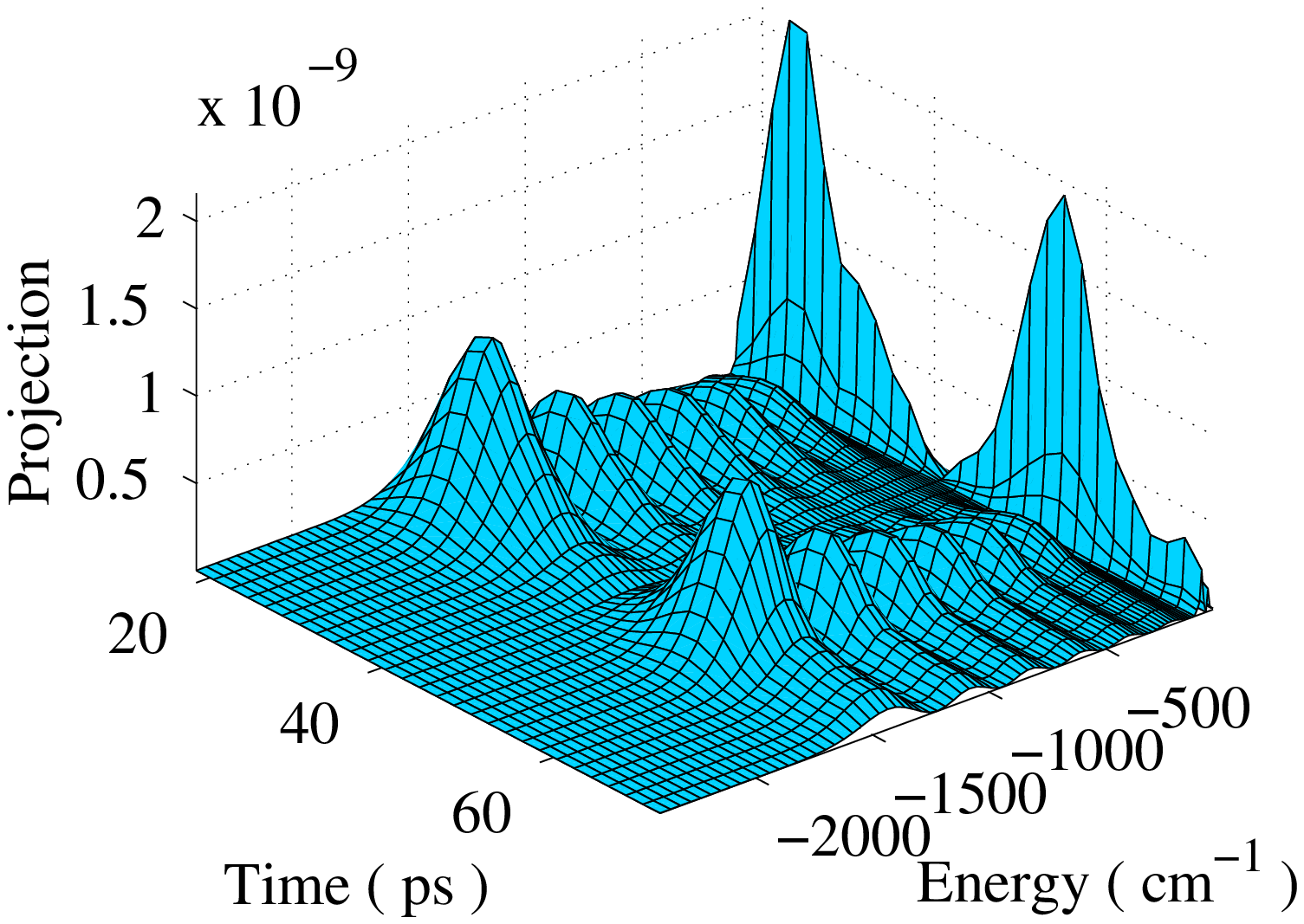}}
\caption{The time-dependence of the projections $P_{\rm
g}^{v^{\prime\prime}}(t)$ of the wavepackets onto the
vibrational levels of the ground electronic state. Left-hand
panel: results obtained with asymptotic spin-orbit coupling
(ASO). Right-hand panel: results obtained with $R$-dependent
spin-orbit coupling (RSO). \label{FCF}}
\end{figure*}

The parameters of the dump pulse are optimized by considering
the overlap of the excited-state wavepacket with the
vibrational levels of the ground electronic state, $P_{\rm
g}^{v^{\prime\prime}}(t)$. Fig.\ \ref{FCF} shows this overlap
as a function of time and binding energy. It may be seen that
the RSO wavepacket has significant overlap with vibrational
levels of the ground electronic state bound by up to about 1500
cm$^{-1}$. The overlap is about 2 orders of magnitude smaller
in the ASO case because of the lack of population near the
inner turning point. This arises from the weaker nonadiabatic
coupling in the ASO case, which in turn arises from the {\em
stronger} spin-orbit coupling near the crossing point. The
time-dependence of the overlap is also rather different in the
ASO case, both because of the larger detuning and because any
population that does arise on the upper adiabatic curve is
trapped there.

\begin{figure}[tbp]
\includegraphics*[width=0.95\linewidth]{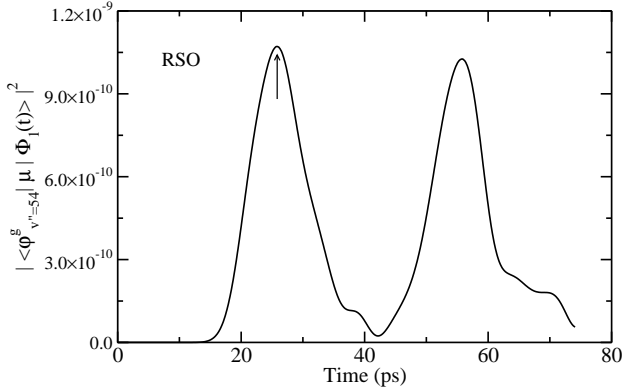}
\caption{The projection of the excited-state wavepacket
onto the $v^{\prime\prime}=54$ vibrational level of the ground
electronic state. The time indicated by the arrow is the maximum
of the dump pulse ($t_{\rm d}$).}
\label{proje2g_54}
\end{figure}

After examining the overlaps for different ground-state
vibrational levels, we chose to de-excite the RSO wavepacket to
level $v^{\prime\prime} = 54$ with binding energy 1435.13
cm$^{-1}$ (shown in Fig.\ \ref{scheme}).
A narrow-bandwidth pulse with temporal width 1 ps,
detuned 1427.61 cm$^{-1}$ to the blue of the pump pulse, is
employed to achieve transfer into this single vibrational level
(because the excited-state wavepacket itself has a binding
energy of 7.5 cm$^{-1}$. As may be seen in Fig.\
\ref{proje2g_54}, the overlap function for this particular
vibrational level has a local maximum at $t \sim 25.8$ ps. The
time delay between the pump and dump pulses is therefore chosen
to give the maximum of the dump pulse at $t_{\rm d} = 25.8$ ps,
corresponding to a time delay of 13.8 ps, as shown in Fig.\
\ref{pulse}. This is in reasonably good agreement with the
classical prediction of 12 ps for a detuning of 7 cm$^{-1}$ in
Table \ref{tab:FCF:RbCs}.

\begin{figure}[tbph]
\includegraphics*[width=0.95\linewidth]{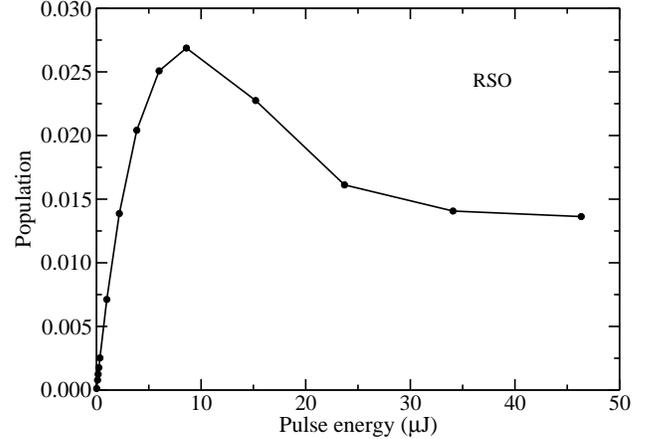}
\caption{Fraction of the excited-state population transferred
to the $v^{\prime\prime}$ = 54 level of the X\,$^1\Sigma^+$
ground state as a function of the dump pulse energy for a
time delay of 13.8 ps.
\label{pop_dump}}
\end{figure}

The final ground-state population $P_{\rm g}^{54}(t)$ is shown
in Fig.\ \ref{pop_dump} as a function of the dump pulse energy.
The renormalisation applied after the pump pulse removes any
population that is already in the ground electronic state
before the dump pulse. Since the dump pulse is resonant only
with the bound $v^{\prime\prime}$ = 54 ground-state level, and
the ground-state vibrational spacing of about 37
cm$^{-1}$ is much larger than the bandwidth of the
pulse, the population of the $v^{\prime\prime}$ = 54 level is
the same as the total ground-state population. The maximum
population transfer is obtained for a pulse of energy $\sim$ 9
$\mu$J. We repeated the dump calculation for the ASO wavepacket
(with a modified laser frequency and a slightly different time
delay to correspond with the maximum in Fig.\ \ref{FCF}) and
verified that the population of ground-state molecules is about
two orders of magnitude smaller.

Ground-state RbCs molecules bound by as much as 1500 cm$^{-1}$
can be formed because there is good Franck-Condon overlap of
the excited-state wavepacket with deeply bound vibrational
levels. Similar studies on the homonuclear Rb$_2$ molecule
\cite{Koch:PRA:2006b,Pechkis:PRA:2007} have shown that excited
Rb$_2$ molecules can be efficiently dumped to create
ground-state vibrational levels bound by only about 10
cm$^{-1}$. For Rb$_2$ the wavepacket showed significant peaking
of singlet population at the {\em outer} tuning point of the
upper adiabatic curve but not at its {\em inner} turning point
as in RbCs. The time delay used for Rb$_2$ was 81.5 ps with a
detuning of 4.1 cm$^{-1}$ from the atomic line.

As commented above, the levels of Rb$_2$ {\em do} show strong
nonadiabatic mixing between the upper and lower adiabatic
states. In view of this, it is important to understand why RbCs
shows much more buildup of singlet character near the inner
turning point than Rb$_2$. The reason for this is that Rb$_2$
has an $R^{-3}$ potential at long range, so that the
vibrational wavefunctions of states near dissociation are more
strongly concentrated at the outer turning point. Because of
this, even population that is transferred to the upper
adiabatic state by nonadiabatic coupling has a smaller
amplitude near the inner turning point and is less effective in
allowing deexcitation to deeply-bound states.

We also note that, because of the small FC factors for
deexcitation, only about 3\% of the excited-state RbCs
population can be transferred to the desired deeply bound
ground-state vibrational level, whereas almost 50\% of the
excited Rb$_2$ population can be transferred to the ground
state using much less dump pulse energy (Fig.\ 11 of
\cite{Koch:PRA:2006b}). The overall efficiency of the scheme
considered here is limited by the fraction of the initial
ground-state population that is transferred to the excited
state. One possible way to enhance it would be by
Feshbach-optimised photoassociation \cite{Pellegrini:2008}.

\section{Conclusions}
We have studied the possibility of using photoassociation with
laser pulses to produce deeply bound RbCs molecules in the
electronic ground state. We have also explored the differences
between photoassociation processes for RbCs and Rb$_2$ in order
to understand the consequences of the different long-range
potentials in excited states of heteronuclear and homonuclear
molecules.

The difference between $R^{-6}$ potentials for heteronuclear
molecules and $R^{-3}$ potentials for homonuclear molecules
produces several important effects. First, the Franck-Condon
factors for excitation are smaller in the heteronuclear case,
though they die off more slowly with detuning from the atomic
line. Secondly, the time taken for a wavepacket produced on the
upper state to evolve from long to short range is significantly
shorter for a heteronuclear molecule than for a homonuclear
molecule of similar mass at the same detuning. We give a simple
semiclassical expression relating the time delay to the
detuning and the coefficients governing the long-range
potential.

We have also explored the dependence of Franck-Condon factors
on the scattering lengths for both the ground state and the
excited state. The Franck-Condon factors oscillate as a
function of detuning, with faster oscillations for RbCs than
for Rb$_2$. Adjusting the ground-state scattering length alters
the positions of the peaks in the Franck-Condon factors as a
function of detuning, while adjusting the excited-state
scattering length leaves the oscillations unchanged but shifts
the vibrational levels within them.

We have carried out wavepacket calculations to explore the
quantum dynamics of RbCs on $0^+$ excited states formed by
coupling the A\,$^1\Sigma^+$ and b\,$^3\Pi$ electronic states.
The diabatic potential curves for the A and b states cross near
$R=10\,a_0$. The dynamics are strongly affected by the
magnitude of the spin-orbit coupling. If the spin-orbit
coupling function is held constant at its asymptotic value, the
dynamics takes place almost independently on the upper and
lower adiabatic curves. Under these circumstances the
wavepacket has almost entirely triplet character at short range
and the Franck-Condon factors for deexcitation to deeply-bound
levels of the ground state are very poor. However, if the
spin-orbit coupling is given a more realistic experimentally
derived form \cite{Fellows:priv:2007} that has a smaller value
near the crossing point, the dynamics are strongly
nonadiabatic. A substantial part of the population is
transferred to the upper adiabatic state, which has singlet
character at short range. This allows efficient deexcitation to
levels of the ground electronic state bound by up to 1500
cm$^{-1}$.

The behavior observed for RbCs may be contrasted with that for
Rb$_2$ \cite{Koch:PRA:2006b, Pechkis:PRA:2007}. Rb$_2$ has only
slightly weaker nonadiabatic coupling but deexcitation is
favored to ground-state levels bound by up to only 10
cm$^{-1}$. The difference in this case arises from the
different long-range potentials: for Rb$_2$ there is an
$R^{-3}$ potential at long range and the density distribution
for the upper-state levels is dominated by the outer turning
point of the upper adiabatic curve. From the outer turning
point, the Franck-Condon factors favor deexcitation to
relatively weakly-bound levels of the ground electronic state.

An important new insight from the present work is that the
combination of strong nonadiabatic coupling with a $1/R^6$
potential produces mixed vibrational levels with significant
singlet density both at long range (which facilitates initial
photoassociation) and near the inner turning point of the
singlet state (which facilitates deexcitation to form deeply
bound ground-state levels). This allows `$R$-transfer' in a way
which does not occur for homonuclear species. This phenomenon
is also expected to be present in other heavy heteronuclear
dimers such as KRb or KCs. It is easily identified
spectroscopically in the level spacings or rotational constants
and will play an important role that is not confined to
photoassociation, for example in stimulated Raman adiabatic
pumping (STIRAP) experiments to form deeply bound states.

Our study of pump-dump photoassociation was motivated by the
search for a coherent scheme for forming deeply bound molecules
where the dump pulse does not destroy the molecules that were
created by the pump pulse. STIRAP is another candidate for
this, and has recently been applied successfully to transfer
near-dissociation Cs$_2$ \cite{Danzl:v73:2008,
Danzl:ground:2008} and KRb \cite{Ospelkaus:2008, Ni:KRb:2008}
molecules to deeply bound states. However, STIRAP requires a
few well-separated levels in order to fulfill the adiabaticity
condition and this requirement is not met by the
quasi-continuum of a MOT. This might be circumvented by
applying STIRAP after a photoassociation pump pulse, in order
to optimize the dump step. In other words the pump pulse
prepares a coherent state which in a two-level picture replaces
the upper level, while the target ground-state vibrational
level plays the role of the lower level. Several realizations
of adiabatic passage in two-level systems exist, such as
Stark-chirped rapid adiabatic passage (SCRAP)
\cite{Yatsenko:1999, Rickes:2000} and retroreflection-induced
bichromatic adiabatic passage (RIBAP) \cite{Yatsenko:2003,
Conde:2005}. The underlying concept of these schemes is to
induce a crossing of the two levels and an adiabatic or
diabatic passage of the induced crossing. In our case the
adiabatic passage would need to proceed on a timescale long
compared to the vibrational motion of the wavepacket (about 40
ps) and short compared to spontaneous emission (about 30 ns).
However, to see whether replacing the picosecond dump pulse by
an adiabatic passage scheme is feasible and whether it allows
the dump efficiency to be increased above the 3\% reported here
is beyond the scope of the present study.

\acknowledgments

We thank Tom Bergeman for sending us his unpublished spin-orbit
coupling results and Eliane Luc-Koenig for her comments on the
manuscript. SG is grateful to the Royal Society for an Incoming
Fellowship and to the CoCoChem project for sponsoring
collaborative visits. JMH is grateful to EPSRC for funding of
the collaborative project QuDipMol under the ESF EUROCORES
Programme EuroQUAM. CPK is supported by the Deutsche
Forschungsgemeinschaft within its Emmy Noether programme.

\bibliography{../../all}
\end{document}